\documentclass[12pt]{iopart}
\usepackage{iopams}
\usepackage{color}
\usepackage {graphicx}

\begin{document}

\title[Expansion of finite-temperature BEC in non-TF regime]{Free-fall expansion of finite-temperature Bose-Einstein condensed gas in the non Thomas-Fermi regime}

\author{M.~Zawada$^{1}$,
        R.~Abdoul$^{1}$,
        J.~Chwede\'nczuk,$^{2}$
        R.~Gartman$^{1}$,
        J.~Szczepkowski,$^{3}$
        \L{}.~Tracewski,$^{4}$
        M.~Witkowski,$^{5}$
                W.~Gawlik$^{6}$}
\address{
$^{1}$Institute of Physics, Nicolaus Copernicus University, Grudzi\c{a}dzka 5, 87-100 Toru\'n, Poland,}
\address{
$^{2}$Institute for Theoretical Physics, Warsaw University, Ho\.za 69, 00-681 Warsaw, Poland,}
\address{
$^{3}$Institute of Physics, Pomeranian Academy, Arciszewskiego 22b, 76-200, S\l{}upsk, Poland,}
\address{
$^{4}$Institute of Experimental Physics, University of Wroclaw, Plac Maksa Borna 9, 50-204 Wroc\l{}aw, Poland,}
\address{
$^{5}$Institute of Physics, University of Opole, Oleska 48, 45-052 Opole, Poland,}
\address{
$^{6}$Institute of Physics, Jagiellonian University, Reymonta 4, 30-057 Krak\'ow, Poland.}

\ead{zawada@fizyka.umk.pl}

\begin{abstract}

We report on our study of the free-fall expansion of a
finite-temperature Bose-Einstein condensed cloud of ${}^{87}$Rb.
The experiments are performed with a variable total
number of atoms while keeping constant the number of atoms in the
condensate. The results provide evidence that the BEC dynamics
depends on the interaction with thermal fraction. In particular,
they provide experimental evidence that thermal cloud
compresses the condensate.

\end{abstract}


\submitto{\JPB}

Anisotropic expansion of a condensate after its release
from a trap is a well known feature of the Bose-Einstein condensed
state \cite{Cor95,Ket95} and is also observed in
non-condensed Bose gases \cite{Wal02,Ger04_t} and in degenerate
Fermi gases \cite{Tho02,Sal03}. At low but finite temperatures,
after switching off the trap potential, free expansion of a
diluted Bose-Einstein condensed gas results in spatial separation
of the thermal and condensed phases \cite{Str99}. This behavior
allows identification of the condensed and thermal fractions
through absorption imaging.

So far, most of the experimental work on the condensed gases was
concentrated on samples with large number of atoms at very low
temperatures, in the so-called Thomas-Fermi (TF) regime. The
number of atoms in the condensed fraction in temperatures close to
$T_C$ is usually small, hence the internal energy of the
condensate fraction is also small compared to its kinetic energy.
On the other hand, in the TF regime, i.e. for condensates with
large number of atoms in temperatures much lower than $T_c$, the
internal energy is large compared to the kinetic one. The
condensate dynamics depends on the number of condensed
atoms: the TF condensates behave hydrodynamically
\cite{Cas96} but for smaller number of condensed atoms,
the effective potential formed from the intrinsic self-interaction
due to ground-state collisions vanishes and the TF approach ceases
to be a good approximation of the BEC \cite{Coo96}. Consequently,
the condensate dynamics can depart from the familiar hydrodynamic
behaviour of TF condensates.

The need for determination and interpretation of this modified
dynamics justifies considerable interest in the region of higher
temperatures, less than, but comparable to $T_C$. Various
properties of finite-temperature Bose-Einstein condensates and
density distributions of their thermal and condensed parts were
studied by several groups both theoretically
\cite{Gio97,Hut97,Wal00,Bac01,Bho02,Jac02} and experimentally. The
experiments on the BEC dynamics were largely devoted to
spectroscopy of collective oscillation modes of the condensate
described in references \cite{Cor97,Ket98,Foo01} which
demonstrated that the mutual interaction of the condensate and
noncondensate components affects the dynamics of the condensate.
In other experiments, Busch et al.\ \cite{Bus00} reported
observation of repulsion of non-condensed gas from the condensate
and Gerbier et al.\ \cite{Ger04} studied the thermodynamics of an
interacting trapped Bose-Einstein gas below $T_C$. With an
assumption that the condensate is in the TF regime, they showed a
temperature dependent deviation from the predicted expansion.

In this report we focus on getting further insight into interaction
between the thermal and condensate fractions of the condensate
which is not in the TF regime. In the first part of our paper we study a pure but small BEC while in the second part we focus on a mixture of a BEC with thermal fraction. Both cases are studied in a free expansion of an atomic sample released from a trap. In the first case with a pure condensate the departure from the
TF predictions is caused by an intrinsic kinetic energy of a small sample,
but in the second case it also reflects interaction of an expanding Bose-Einstein
condensed cloud with thermal atoms.
Our measurements show that the behavior of the condensate part
depends on both the number of condensed atoms, $N_{0}$, and on its
interactions with thermal component. Thus, to study the effect of
BEC interaction with thermal cloud, the BEC atom number
needs to be kept constant when varying the $N_{0}/N$ ratio ($N$
being the total atom number). These measurements confirm that the
BEC dynamics strongly depends on interaction with thermal fraction
and provide clear evidence of the compression effect.

Our experimental setup uses magnetic trap with longitudinal and
axial frequencies, $\omega_z/2\pi = 12.(07) \pm 0.38$~Hz and $\omega_r/2\pi =
13(7.4) \pm 5.8$~Hz, respectively. We create BEC of up to 300~000 ${}^{87}$Rb
atoms in the $\vert F=2,m_F=2\rangle$ hfs component of their
ground state. This sample is analyzed by absorptive imaging with
the imaging beam resonant with the $\vert F=2,
m_F=2\rangle$~-~$\vert F=3, m_F=3\rangle$ hfs component of the
${}^{87}$Rb D2 line. More experimental details can be found in
Ref.\ \cite{Byl08}.

All pictures are taken after releasing atoms from the MT with a
delay which can be varied between 2 and 30~ms in our setup. The
images of a finite-temperature condensates reveal two
fractions, the BEC and the thermal fraction. Their accurate
identification is essential for proper analysis of the
experimental results \cite{Ket99}. Our method of such
identification will be described elsewhere \cite{Szc08}.

\begin{figure}[hbt]
\begin{center}
      \resizebox{0.6\columnwidth}{!}{\includegraphics{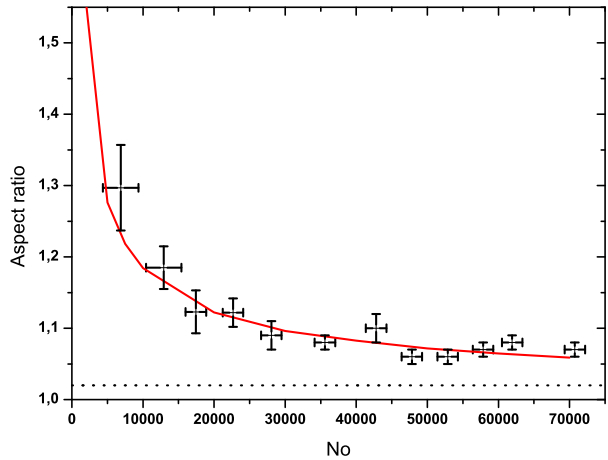}}
      \caption{Aspect ratio of a pure BEC as
      a function of the number of condensed atoms, $N_{0}$. The measurements are performed after 15~ms of the condensate
      free expansion after its releasing from the trap.
The solid (red) line shows the numerical prediction of the 3D
Gross-Pitaevskii equation for a pure BEC. The dotted line is the BEC AR predicted
by the Castin-Dum equation \cite{Cas96}, valid for the TF regime.}
      \label{fig:No_AR}
\end{center}
\end{figure}

\begin{figure}[hbt]
\begin{center}
      \resizebox{0.6\columnwidth}{!}{\includegraphics{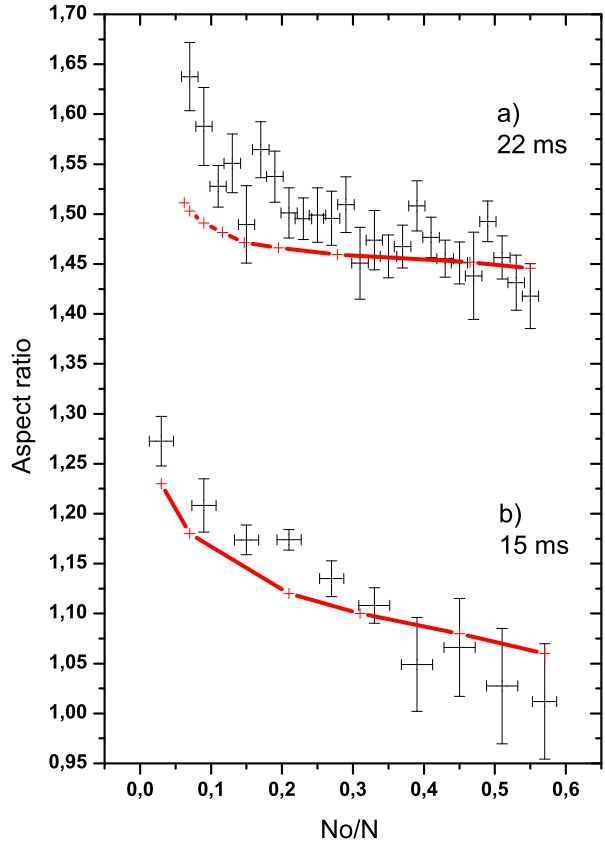}}
      \caption{
Aspect ratio of a finite-temperature BEC vs. condensate fraction, $N_{o}/N$. The AR measurements are made after a) 22 ms and b) 15 ms of free expansion. Solid (red) lines show the numerical predictions for pure BEC behavior, based on the 3D Gross-Pitaevskii equation as a function of $N_{0}$}.

      \label{fig:NoN_AR}
\end{center}
\end{figure}

We start by checking the applicability of the TF approximation and
analyzing the AR($N_{0}$) dependence with a pure condensate. In
case of pure BEC released from the trap, its AR measurements give
information about the ratio between the kinetic and mean-field
energies of the condensate. Such measurements were performed after
15~ms of free expansion and their results are depicted in Fig.\
\ref{fig:No_AR}. The dotted line marks the AR value
obtained by solving the Castin-Dum equations
\cite{Cas96}, valid for the TF regime. During free fall, the shape of our condensate
changes from a cigar-shaped distribution to an oblate
distribution. After 15~ms expansion, the aspect ratio
becomes about 1 but then grows with the expansion time, as can be
seen in Fig.\ \ref{fig:NoN_AR}.
The measurements prove that TF approximation is not applicable to our
BEC, the discrepancy being caused by intrinsic kinetic energy of a small BEC. We model thus
its dynamics by numerically solving of the 3D Gross-Pitaevskii
equation using a split-step method which yields the solid line in
Fig.\ \ref{fig:No_AR}. As seen, for decreasing number of atoms
$N_0$, our experimental results deviate more and more from the TF
model predictions, whereas the 3D GP model describes our
observations very well.

After testing the TF approximation, we present the measurement
results for thermal condensates, i.e. for BECs in the presence of
thermal fractions. Our imaging technique limits the study to the
condensate fractions up to $N_{0}/N=0.6$.
For these measurements, the finite-temperature BEC was released
from the trap and the AR measurements were performed after two
different times of free expansion, 22~ms and 15~ms. Fig.\
\ref{fig:NoN_AR} depicts the AR dependence on the condensate fraction $N_{0}/N$ along with the  numerically simulated AR (solid line) as a function of $N_{0}$, i.e. ignoring the thermal fraction \cite{No}.

There is a noticeable difference between the numerical predictions
ignoring the thermal fraction (solid line in Fig.\
\ref{fig:NoN_AR}) and the measurement results, especially for the
range of small condensate fractions (big thermal fractions). This
deviation is more significant for the longer free fall time
(22~ms) which illustrates importance of the interaction between
the condensate and thermal cloud.

Below we present results of the AR measurement with
thermal condensates but in
contrast to the measurements depicted in Fig.\ \ref{fig:NoN_AR}, we fix the number
of atoms in the condensate phase ($N_{0}$). The values of $N_{0}$ are chosen
to be high enough that the condensate AR does not change with the
small variations of $N_{0}$ (see Fig.\ \ref{fig:No_AR}).

\begin{figure}[hbt]
\begin{center}
      \resizebox{0.6\columnwidth}{!}{\includegraphics{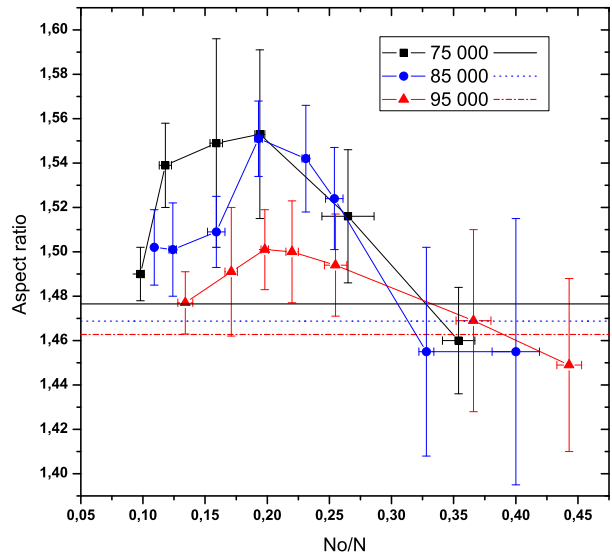}}
      \caption{Aspect ratio of a finite-temperature BEC with constant $N_0$ vs. the condensate fraction, $N_{0}/N$, measured after 22 ms of free expansion. Squares (black), circles (blue) and triangles (red) correspond to three sets of measurements, with $N_0$=75000, 85000 and 95000 atoms, respectively. Solid lines are shown to guide the eye. Horizontal lines depict the numerical 3D GP predictions for pure BECs.

}
      \label{fig:No_const}
\end{center}
\end{figure}

The aspect ratio measurements are made after 22 ms of the free
expansion. Fig.\ \ref{fig:No_const} depicts the AR dependence on
the condensate fraction ($N_{0}/N$) with $N_{0}$=const, so that AR is
analyzed as a function of the number of the thermal atoms only.
Squares, circles, and triangles correspond to three sets of
measurements, with $N_{0}$=75000, 85000 and 95000 atoms, respectively.
Solid lines are shown to guide the eye. Dotted and dashed lines
show the numerical prediction for pure BEC behavior, based on the
3D Gross-Pitaevskii equation solved with a split-step method.

\begin{figure}[hbt]
\begin{center}
      \resizebox{0.6\columnwidth}{!}{\includegraphics{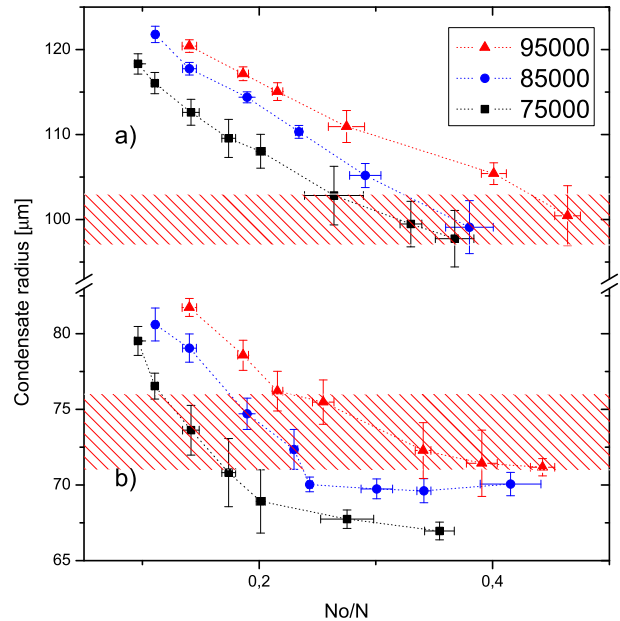}}
      \caption{The radial (a) and axial (b) radii of
      the condensate vs. the condensate fraction ($N_{0}/N$).
      The number of atoms in the condensate phase ($N_{0}$) remains constant
      in the experiment. Squares (black), circles (blue) and triangles (red)
      correspond to three sets of measurements, with $N_0$=75000, 85000
      and 95000 atoms, respectively. Dotted lines are shown to guide the eye. 
Dashed horizontal bands represent radii of a pure BEC with $N_0$=95000 atoms. }
      \label{fig:R_Noconst}
\end{center}
\end{figure}

The deviation from the GP predictions, seen already in
Fig.\ref{fig:NoN_AR}, reflects a mutual interaction of the
condensate and noncondensate components. The effect
becomes even more pronounced in the condensate
size. Fig.\ \ref{fig:R_Noconst} depicts the condensate
radii plotted as a function of the condensate fraction
for the same three sets of the $N_0$ values as in Fig.\
\ref{fig:No_const}.
These results clearly prove that thermal cloud compresses
the BEC. At equilibrium, the shell of thermal atoms surrounding
the condensate in a trap exerts a force toward the trap center,
thereby compressing it. In the free fall this compression results
in a faster expansion in all directions, which is well evidenced
in Fig.\ \ref{fig:R_Noconst}. The anisotropy of the expansion in
the axial and radial directions explains the
non-monotonic behavior of the aspect ratio depicted in
Fig.\ \ref{fig:No_const}.

It is interesting to compare these radii with those of a
pure BEC. For this sake, we depict with dashed bands (the band
widths represent the measurement uncertainties) the axial and
radial radii of a pure condensate of $N_{0}$=95000 atoms, measured
in the same conditions. The corresponding TF radii for
$N_{0}$=95000 and expansion time of 22~ms are 65 $\mu$m (radial)
and 47 $\mu$m (axial), i.e. below the measured values which is to
be attributed to the departure of our condensate from the TF
regime. It would be interesting to extend these measurements for
systematic studies of the energy transfer between thermal and
condensed atoms.

In conclusion, we have observed that the dynamics of the finite
temperature condensates depends on the number of condensed atoms,
$N_{0}$, and on the ratio between $N_{0}$ and total number of
atoms, $N_{0}/N$, which reflects effects of the
interaction of BEC with non-condensed thermal atoms. The main
result of this work is the clear experimental evidence that the
thermal cloud compresses BEC.

\section*{Acknowledgments}

This work has been performed in KL~FAMO, the National Laboratory of AMO Physics in Toru\'n and supported by the Polish Ministry of Science. The authors are grateful to  M. Gajda, M. Brewczyk, K. Gawryluk and J. Zachorowski for numerous discussions. J.S. acknowledges also partial support of~the~Pomeranian University (projects numbers
BW/8/1230/08 and BW/8/1295/08).


\begin{thebibliography}{10}
\bibitem{No} For a given fraction $N_{0}/N$, $N_0$ itself varies,
so in the numerical simulation we used an average value of $N_0$ for each value of $N_{0}/N$ fraction.
\end{thebibliography}
\end{document}